\documentclass{article} 
\usepackage{nips13submit_e,times}
\usepackage{hyperref}
\usepackage{url}
\usepackage{graphicx}
\usepackage{float}
\usepackage{pbox}
\usepackage{caption}
\usepackage{cite}
\usepackage{amsmath}
\usepackage{amssymb}
\usepackage{enumerate}
\usepackage{tablefootnote}

\DeclareMathOperator*{\argmax}{\arg\!\max}

\title{Diagnosing ADHD from fMRI Scans Using \\ Hidden Markov Models}

\author{
Bhaskar Sen, Zheng Shi, and Gregory Burlet \\
Department of Computer Science \\
University of Alberta \\
Edmonton, AB T6G 2E8 \\
\texttt{\{bsen, zshi3, gburlet\}@ualberta.ca}
}

\nipsfinalcopy 

\begin{document}

\maketitle

\begin{abstract}

This paper applies a hidden Markov model to the problem of Attention Deficit Hyperactivity Disorder (ADHD) diagnosis from resting-state functional Magnetic Resonance Image (fMRI) scans of subjects. The proposed model considers the temporal evolution of fMRI voxel activations in the \emph{cortex}, \emph{cingulate gyrus}, and \emph{thalamus} regions of the brain in order to make a diagnosis. Four feature dimensionality reduction methods are applied to the fMRI scan: voxel means, voxel weighted means, principal components analysis, and kernel principal components analysis. Using principal components analysis and kernel principal components analysis for dimensionality reduction, the proposed algorithm yielded an accuracy of 63.01\% and 62.06\%, respectively, on the ADHD-200 competition dataset when differentiating between healthy control, ADHD innattentive, and ADHD combined types.

\end{abstract}

\section{Introduction}

Statistical machine learning methods have recently permeated disciplines such as Psychiatry, which specialize in the diagnosis and treatment of neuropsychiatric disorders. The availability of large-scale functional Magnetic Resonance Image (fMRI) datasets have encouraged the application of advanced machine learning models to the diagnosis of neuropsychiatric disorders \cite{cecchi2009}. fMRI scans measure brain activity by detecting fluctuations in blood-oxygen levels over time. Brain activations are represented digitally as \emph{voxels}, the three-dimensional analogue of pixels.

In this paper we present the application of a temporal model, specifically a Hidden Markov Model (HMM), to the problem of automatically diagnosing Attention Deficit Hyperactivity Disorder (ADHD) from resting-state fMRI scans of subjects.\footnote{A \emph{resting-state} fMRI measures the brain activities of subjects that are not asked to perform a given task during the course of the scan.} ADHD is a psychiatric disorder that adversely affects the attention span, hyperactivity, or impulsivity of an individual. Since ADHD positive individuals have difficulty maintaining focus on a mental activity, it is reasonable to assume that the temporal evolution of their brain activities, as measured by the fMRI, differ from those of healthy individuals. It is with this argument that we motivate the use of a temporal model for ADHD diagnosis.

The primary challenge in developing a machine learning algorithm for ADHD diagnosis is the large dimensionality of the fMRI scan. A single fMRI scan may consist of hundreds of three-dimensional images over time, each of which is composed of approximately 500,000 voxels. Therefore, extracting lower dimensional fMRI representations that retain discriminative features for diagnosis is an important step in the implementation of diagnosis algorithms. If successful, an automatic ADHD diagnosis algorithm would aid mental-health care professionals in the diagnosis and treatment of the disorder.

This paper is structured as follows: the following section provides a review of related work. Section~\ref{sec:algorithm} describes the proposed temporal model for ADHD diagnosis and outlines the feature extraction and feature representation techniques explored. Section~\ref{sec:evaluation} describes the dataset used for training and evaluating the proposed algorithm. In addition, this section states hypotheses regarding the performance and structure of the learned diagnostic model and describes the evaluation procedure for testing the performance of the algorithm. The experimental results and analysis are presented in Section~\ref{sec:results}. Finally, the work is concluded in Section~\ref{sec:conclusion}.

\section{Related Work}
\label{sec:relatedwork}


Using the ADHD-200 competition dataset, which consists of several hundred resting-state fMRI scans,\footnote{\url{http://fcon_1000.projects.nitrc.org/indi/adhd200}} Eloyan \emph{et al.} \cite{caffo2012} explored several different classifiers for ADHD diagnosis, including a support vector machine, gradient boosting, and voxel-based morphology. In addition, several feature extraction methods were investigated, including singular value decomposition and CUR matrix decomposition. The best classification accuracy was achieved by taking a weighted combination of these classifiers, which yielded $61.0\%$ accuracy on the test data. Also using the ADHD-200 competition dataset, Sina \emph{et al.} \cite{Sina2013} extracted histogram of oriented gradient features from fMRI scans, which were then input to a support vector machine. The classifier yielded an accuracy of $62.6\%$ on the test dataset. These two methods report the highest classification accuracy on the ADHD-200 competition dataset.

Recent studies \cite{bullmore2009,greicius2002} emphasize that different parts of the brain are functionally correlated. Taking this into consideration, the Human Connectome Project explores graphical models that seek to capture these functional connectivities, both in task-based and resting-state fMRI scans.\footnote{\url{http://humanconnectome.org/}} Similarly, Zhang \emph{et al.}\cite{Zhang2005} proposed a Bayesian network for modeling functional neural activity. In this work, each region of the brain is represented as a node in the graphical model and the functional connectivity of these nodes over time is used for classifying drug addicts from healthy controls.

Apart from using functional relations, temporal correlation between brain voxels and connectivity has been explored by Fiecas \emph{et al.} \cite{fiecas2013}. Furthermore, the temporal relation between mental states and neuronal activities has been investigated by building a conditional random field \cite{Li2011}. Duan \emph{et al.} \cite{Duan2005} proposed two methods based on likelihood and distance measures to analyze fMRI scans using an HMM. However, this work focuses on analyzing the Blood Oxygen Level Dependent (BOLD) signal for brain activities in order to predict the brain activations in task-based fMRI time series. Eavani \emph{et al.} \cite{Eavani2013} analyzed the functional connectivity dynamics in resting state fMRI and decoded the temporal variation of functional connectivity into a sequence of hidden states using an HMM. 

Similar to the previously mentioned temporal approaches to ADHD classification, we will investigate the temporal evoluation of voxels for both healthy and ADHD positive subjects using an HMM. However, we explore reduced dimensional representations of fMRI voxels in ADHD regions of interest. The analysis of regions of interest in an fMRI scan instead of the entire brain is common practice. For instance, Solmaz \emph{et al.} \cite{solmaz2012} used a bag of words approach for identifying ADHD patients using the \emph{default mode network} region of the brain.

\section{Temporal ADHD Diagnosis Algorithm}
\label{sec:algorithm}

\begin{figure}[ht]
	\centering
	\includegraphics[width=0.8\textwidth]{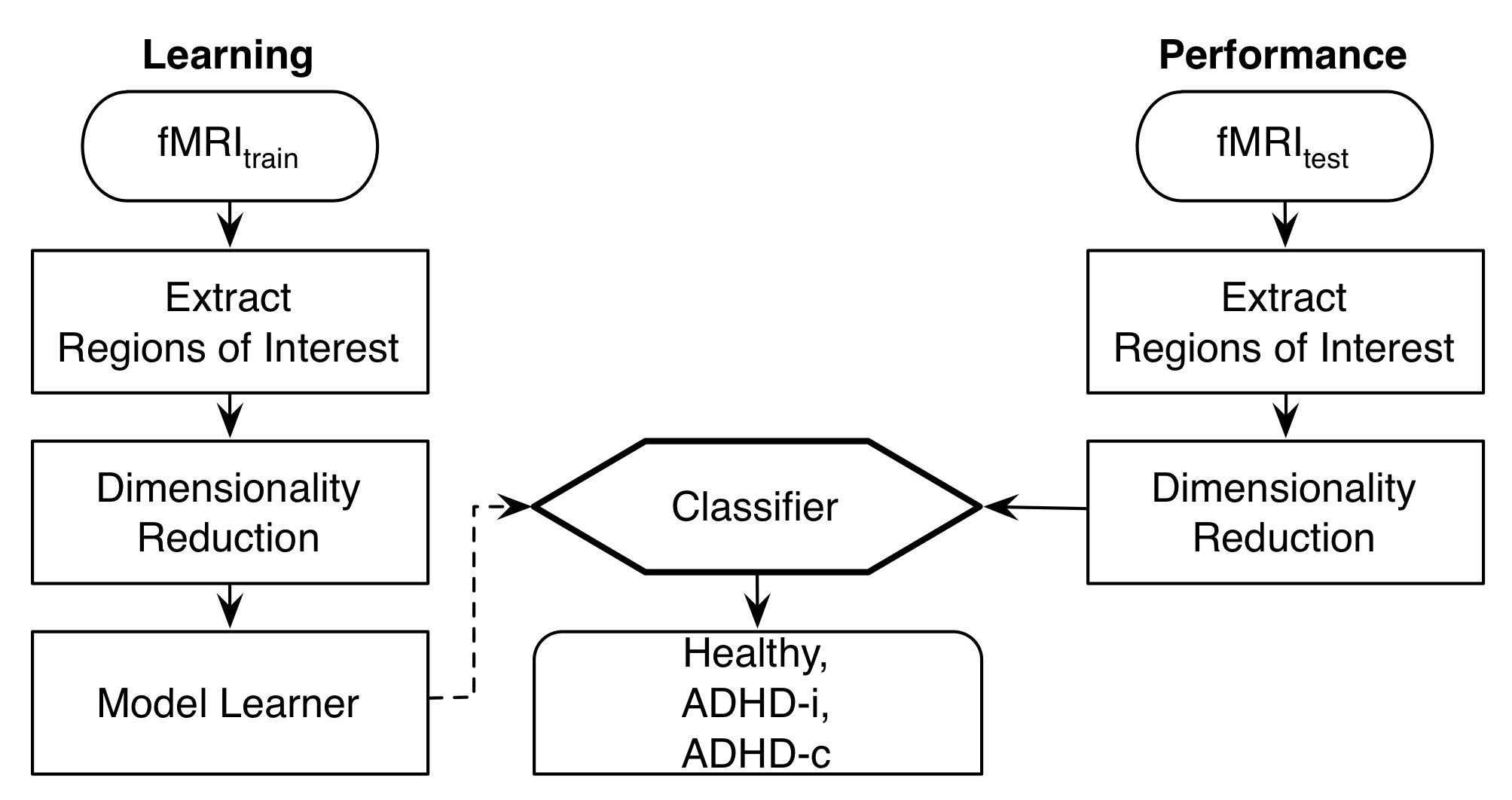}
    \caption{Input, algorithm pipeline, and output of the learning (left) and performance task (right) of our algorithm.}
    \label{fig:algorithmio}
\end{figure}

In order to learn an ADHD classification model, several steps are necessary. For each time slice of the fMRI, voxels are extracted from ADHD Regions of Interest (ROIs) in the brain and dimensionality reduction algorithms are applied to the voxels in each region to reduce the dimensionality of the data. The resulting data is presented as observations to a cluster of Hidden Markov Models (HMMs) that learn to discriminate between healthy, ADHD inattentive, and ADHD combined types. Using the resulting classifier, new fMRI data can be input to the system; features are extracted from the fMRI data, which the classifier uses to diagnose the subject. This process is depicted in Figure~\ref{fig:algorithmio}. Each step of the process is explained in more detail in the following sections.

\subsection{Extracting ADHD Regions of Interest}

For each time slice of each fMRI scan, we extract clusters of voxels that correspond to the twelve ADHD regions of interest from the cortex, cingulate gyrus, and thalamus regions of the brain \cite{bush2005}, according to the Harvard-Oxford Cortical and Subcortical Structural Atlas.\footnote{\url{http://neuro.debian.net/pkgs/fsl-harvard-oxford-atlases.html}} Each ROI is composed of approximately 4000 voxels; there are 48,710 voxels across all ROIs, which is less than a tenth of the $\approx 500,000$ voxels in one time slice of an fMRI scan. Despite this reduction in dimensionality, the dimensionality of the feature space is still quite large. To further reduce the dimensionality of the feature space, four dimensionality reduction algorithms---described in the following sections---are applied to the fMRI voxels extracted from the twelve ADHD regions of interest.

\subsection{Feature Dimensionality Reduction}

\subsubsection{Voxel Means}
\label{subsec:ROImeans}

An unsophisticated method of dimensionality reduction is to simply compute the average value of the given feature set. Using this method, we compute the average of the fMRI voxel values in each region of interest. The output of this dimensionality reduction method is a matrix $O \in \mathbb{R}^{m \times t \times 12}$, such that $m$ is the number of subject fMRI scans in the dataset, $t$ is the number of image samples over time, and 12 is the number of ROIs.

\subsubsection{Weighted Voxel Means}
\label{subsec:ROIweightedmeans}
The average value of the voxels ignores the fact that the voxels lying near the centre of a region of interest may contain more information than the voxels at the boundary. Taking this into consideration, we experimented with computing a weighted mean of voxels for each region of interest. The weights were derived from a univariate Gaussian distribution with a standard deviation of one and a mean value that is aligned with the center of that region. Hence, voxels further from the centre will be assigned smaller weights. The output of this dimensionality reduction method is a matrix $O \in \mathbb{R}^{m \times t \times 12}$, such that $m$ is the number of subject fMRI scans in the dataset, $t$ is the number of image samples over time, and 12 is the number of ROIs.

\subsubsection{Principal Components Analysis}
\label{subsec:ROIPCA}

Principal Components Analysis (PCA)\cite{joll2002} applies an orthogonal transformation to an input dataset to convert a set of possibly correlated variables into a set of linearly uncorrelated variables named principal components. PCA is applied to the voxels in each region of interest and the three principal components with the largest spectral components are selected. The term \emph{spectral components} refers to the singular values in the singular value decomposition of the input data matrix. Three principal components were selected because these singular values were substantially larger than the others, and thus, these principal components capture the most information about the input voxel data. The output of this dimensionality reduction method is a matrix $O \in \mathbb{R}^{m \times t \times 36}$, such that $m$ is the number of subject fMRI scans in the dataset, $t$ is the number of image samples over time, and $3 \times 12 = 36$ is a concatenated vector containing the three principal components computed for each of the twelve regions of interest.

\subsubsection{Kernel Principal Components Analysis}
\label{subsec:kPCA}

Kernel Principal Components Analysis (kPCA) is a non-linear dimensionality reduction technique that maps the data to a non-linear feature space that is defined by a kernel function. The mapping attempts to unfold the data onto a lower dimensional manifold. In the context of our project, we assume that the fMRI voxels in each region of interest lie on a ten dimensional manifold, which the kPCA algorithm will attempt to recover. Given an fMRI scan, for each time instance we apply kPCA on each of the twelve regions of interest. In order to stretch the underlying lower dimensional manifold, the feature space should maximize the distance between two neighboring points while keeping the locality constraint intact. Formally, if $\mathbf{x}_{i}$, $\mathbf{x}_{j}$, and $\mathbf{x}_{k}$ are three neighboring points and their corresponding feature-space representation is $\Phi_{i}, \Phi_{j}$, and $\Phi_{k}$ then the problem of manifold learning becomes
\begin{equation}
\max_{\mathbf{\phi}} \|\Phi(\mathbf{x}_i)-\Phi(\mathbf{x}_j)\|_F^2,
\end{equation}
subject to the constraint that $(\Phi(\mathbf{x}_{i})-\Phi(\mathbf{x}_{j}))^T(\Phi(\mathbf{x}_{i})-\Phi(\mathbf{x}_{k}))=(\mathbf{x}_{i}-\mathbf{x}_{j})^T(\mathbf{x}_{i}-\mathbf{x}_{k})$ and $\sum_{i}\Phi(\mathbf{x}_{i}) = 0$ \cite{Sha2004}. This optimization problem can be efficiently solved by semidefinite programming techniques. The output of this dimensionality reduction method is a matrix $O \in \mathbb{R}^{m \times t \times 120}$, such that $m$ is the number of subject fMRI scans in the dataset, $t$ is the number of image samples over time, and $10 \times 12 = 120$ is a concatenated vector containing the ten dimensional subspace computed by kPCA for each of the twelve regions of interest.


\subsection{Hidden Markov Model Classifier}

\begin{figure}[ht]
	\centering
	\includegraphics[width=0.8\textwidth]{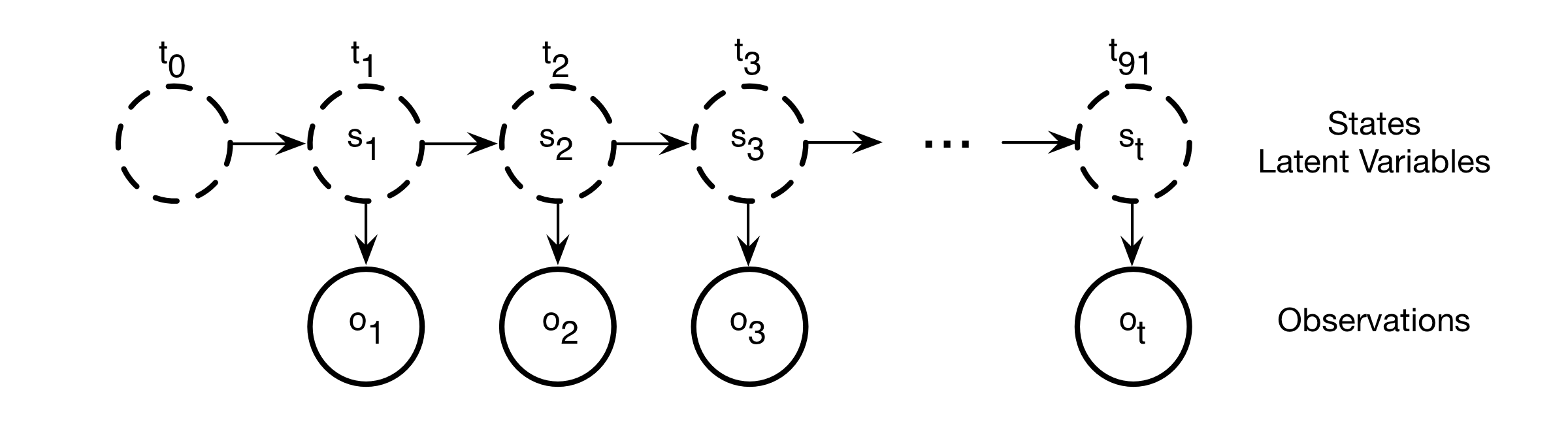}
    \caption{Structure of the hidden Markov model for ADHD classification. Observations of the model are the different feature representations of the voxels in the regions of interest for ADHD. States of the model correspond to the current \emph{mental state} of the brain.}
    \label{fig:hmmstructure}
\end{figure}

Regardless of the dimensionality reduction algorithm used, the feature output is a matrix $O \in \mathbb{R}^{m \times t \times n}$, such that $m$ is the number of subject fMRI scans, $t$ is the number of image samples over time, and $n$ is the size of the feature set after dimensionality reduction. The sequence of fMRI features of a subject are posed as \emph{observations} to an HMM, a probablistic graphical model with the structure presented in Figure~\ref{fig:hmmstructure}. The latent variables (hidden states) of the HMM correspond to the current \emph{mental state} of the brain. For example, the mental state of the brain can be interpreted as the task or thought being focused on at any given time.

For training the model, the fMRI scans in the dataset described in Section~\ref{subsec:dataset} are partitioned into groups according to their corresponding class label: healthy (1), ADHD inattentive (2), or ADHD combined (3). One HMM is trained for each class label. Each HMM $\lambda_i, \forall i \in \{1,2,3\}$ is initialized with random values for the initial state distribution, the transition matrix, and the emission distribution. The emission distribution is a Gaussian mixture model. Preliminary experiments using mean voxel values for regions revealed that a mixture of five Gaussian distributions yielded optimal results. From the dataset, the model parameters of each HMM were learned using the Baum-Welch algorithm \cite{rabiner1989}. The Probabilistic Modeling Toolbox for Matlab was used for model initialization, training, and classification.\footnote{\url{http://github.com/probml/pmtk3}}

Given an fMRI scan of a single subject $O \in \mathbb{R}^{t \times n}$, ADHD classification is performed by returning the model $\lambda_i$ that maximizes the probabililty of the observations $O$
\begin{equation}
\label{eqn:classification}
\argmax_{i\in\{1,2,3\}} P(O | \lambda_i),
\end{equation}
where $P(O | \lambda_i)$ is computed by summing the forward variables in the \emph{Forward-Backward} Procedure, which Rabiner \cite{rabiner1989} meticulously describes.

\section{Algorithm Evaluation}
\label{sec:evaluation}

This section will describe the fMRI dataset used to train and evaluate the proposed temporal ADHD diagnosis model. Several hypotheses are made regarding the performance and structure of the model. To confirm or refute these conjectures, several experiments are proposed.

\subsection{Dataset}
\label{subsec:dataset}

The ADHD-200 competition dataset is used to train and evaluate our ADHD diagnosis model. The dataset consists of 940 fMRI scans of subjects that are labeled by health-care professionals as healthy control, ADHD inattentive, ADHD impulsive, or ADHD combined. Only 39 fMRI scans are labeled as ADHD impulsive, so these scans were removed from the dataset. There are 91 image samples over time for each subject fMRI scan. Therefore, the dataset is of the form $O \in \mathbb{R}^{901 \times 91 \times n}$, where the raw fMRI scans contain $n=510,340$ voxels per time sample.

The ADHD-200 competition dataset preprocesses the raw fMRI scans using the following transformations: motion correction, which compensates for patient head movement; co-registration, which projects each time slice of an fMRI scan to a standard structural MRI scan space; normalization, which transforms the size and shape of each fMRI scan to have the same dimensions; temporal filtering, which removes drift or noise from the time-series data; and intensity normalization, which normalizes the voxel values to lie within the range $[0,2]$.

\subsection{Hypotheses}

We propose several hypotheses regarding our diagnosis model:
\begin{enumerate}[(i)]

\item Increasing the model complexity (number of hidden states in the HMM) will increase classification accuracy.

\item In terms of the dimensionality reduction algorithm used, the classification accuracy of the proposed diagnosis system using Kernel PCA features will be greater than PCA, which will be greater than weighted voxel means, which in turn will be greater than voxel means. The rationale for this hypothesis is that taking the mean or weighted mean value of voxels in each of the twelve ROIs na\"{i}vely reduces the dimensionality of the feature space such that discriminative features for proper classification are discarded. We hypothesize that the alternative dimensionality reduction algorithms will preserve discriminative features.

\item After training, the trace of the state transition matrix for the healthy HMM $\lambda_1$ is higher than the trace of the state transition matrices for the ADHD-i HMM $\lambda_2$ and the ADHD-c HMM $\lambda_3$. Recall that the trace operator of a matrix sums the diagonal entries of the matrix, which are the probabilities that the mental state of the subject stays the same. The rationalization for this conjecture is that healthy subjects will tend to stay in the same mental state in contrast to subjects that are ADHD positive.

\end{enumerate}

\subsection{Evaluation Method}
We use 5-fold cross validation to evaluate our implemented diagnosis algorithm. The fMRI dataset is partitioned into five subsets, such that in each iteration 4/5 of the dataset is used for training and 1/5 of the dataset is used for testing. A different subset is used for testing in each iteration. Care is taken to partition the dataset such that each subset is populated with an equal distribution of class labels, i.e., each of the five subsets contain an equal proportion of fMRI data corresponding to healthy, ADHD-i, and ADHD-c types. When testing on the $k^{\text{th}}$ subset of data, our classifier will output a class label $\mathbf{\hat{y}} \in \{\text{healthy, ADHD-i, ADHD-c}\}$ for each of the $m_k$ fMRI scans. Given the ground-truth labels $\mathbf{y}$, the accuracy of our classifier on fold $k$ is 
\begin{equation}
\label{eqn:foldacc}
{\tt accuracy}(k) = \frac{\sum_{i=1}^{m_k} I{\{\mathbf{y}_i = \mathbf{\hat{y}}_i\}}}{m_k}.
\end{equation}
The final classification accuracy is reported as the average of the classification accuracies computed for each of the five cross-validation folds (\ref{eqn:foldacc}).

To confirm or refute the hypotheses posed in the previous section, several experiments have been designed. To test hypothesis (i--ii), for each dimensionality reduction method proposed, 5-fold cross validation will be conducted for the following number of HMM states $[4, 8, 12, 16, 20]$. In this case, each of the three HMMs will have the same state value. Hypothesis (ii) can be accepted or rejected based on the results reported by the experiment proposed for hypothesis (i). To test hypothesis (iii), using the dimensionality reduction method and number of states parameter that received the highest accuracy in experiment (i), this model will be trained on the entire dataset and the trace of the transition matrix for the healthy HMM will be compared to the trace of the transition matrix for the ADHD-i and ADHD-c HMMs.

\section{Results}
\label{sec:results}

The results of the experiment proposed for hypothesis (i) and (ii) are presented in Table~\ref{tab:cvresults}, which displays the 5-fold cross-validation accuracy for different numbers of hidden states in the HMM as well as different fMRI dimensionality reduction methods.


Our first hypothesis states that classification accuracy will increase as the model complexity increases. Analyzing the 5-fold cross-validation accuracy of the proposed model for different numbers of hidden states in the HMMs, (Table~\ref{tab:cvresults}) reveals that our first hypothesis is not quite correct. In the case of the voxel mean, weighted voxel mean, and kernel PCA dimensionality reduction methods, we see an increase in cross-validation accuracy as the model complexity increases until a point where further increases in complexity actually hinders classification performance. In the case of PCA, we see minute fluctuations in the classification accuracy as model complexity increases; however there is no apparent trend. This phenemenon suggests that when processing PCA features of fMRI scans, classification of ADHD may be independent of the internal hidden state of the HMM and alternative temporal models should be explored in this case.

For each dimensionality reduction method, if we consider the number of states parameter that yields the maximum cross-validation classification accuracy (typeset in bold in Table~\ref{tab:cvresults}), our hypothesis that the classification accuracy of kernel PCA will exceed PCA, which will exceed weighted voxel means, which will exceed the accuracy of voxel means, is almost correct. The results show that the classification accuracy of PCA and kPCA dimensionality reduction is indeed greater than weighted voxel means, which is greater than voxel means. However, the accuracy of PCA and kPCA are almost equivalent and outperform state-of-the-art diagnosis systems on the ADHD-200 dataset \cite{caffo2012,Sina2013}. Hence, we conclude that for PCA, the three components along the maximum variability of data successfully captures a significant amount of information necessary for classifying ADHD. Also, our assumption that the fMRI data for each subject and for each time instance lies on a lower dimensional manifold is also true for the purposes of classification.

Our third hypothesis states that when training our classifier with 20 hidden states for each of the three HMMs on the entire fMRI dataset using PCA dimensionality reduction, the trace of the transition matrix for the healthy HMM will be larger than the trace of the transition matrices for the ADHD-i and ADHD-c HMMs. The results of this experiment are presented in Table~\ref{tab:transtrace}. Though the results show that our hypothesis is in fact correct, the differences in trace are too insignificant to draw any concrete conclusions. Perhaps the most interesting question that arises from this experiment is why the trace of the transition matrices are so small. Note that a trace value of two on a transition matrix of the form $A \in [0,1]^{20 \times 20}$ implies that each self-state transition in the HMM is approximately $2/20 = 0.1$. This suggests that most of the transitions among the underlying Markov chain may be from one state to another instead of from one state to itself. 

Upon further investigation of the transition matrices, it seems the opposite is true. In all cases, i.e., for all numbers of hidden states, and for all dimensionality reduction algorithms explored, one state in the Markov chain has a high self-state transition probability in the order of approximately 0.99. That is, 99\% of the time, the Markov chain will remain in this state once it wanders onto the state. This means that a single state is essentially attempting to explain every single time slice of the fMRI, which is not the desired effect of using such a temporal model.

\setcounter{footnote}{0}
\renewcommand{\thefootnote}{\fnsymbol{footnote}}

\begin{table}[t]
\caption{5-fold cross-validation results for ADHD classification (healthy, ADHD-i, or ADHD-c) using various numbers of HMM states for the following feature representations: voxel means, weighted voxel means, PCA, kPCA}
\label{tab:cvresults}
\begin{center}
\begin{sc}
\begin{tabular}{c c c c c}
\hline
Number of States & Voxel Means & \pbox{20cm}{$\;\,$ Weighted \\ Voxel Means} & PCA\tablefootnote{Using 9 Gaussian mixtures instead of 5.} & kPCA\tablefootnote{Using 16 Gaussian mixtures instead of 5.} \\
\hline
4 & 44.86\% & 48.55\% & 62.80\% & 60.22\% \\
8 & 49.80\% & \textbf{54.16\%} & 62.26\% & 61.07\% \\
12 & 51.58\% & 52.34\% & 61.84\% & 61.50\% \\
16 & \textbf{53.89}\% & 49.27\% & 62.15\% & \textbf{62.06}\% \\
20 & 48.49\% & 47.64\% & \textbf{63.01}\% & 60.62\% \\
\hline
\end{tabular}
\end{sc}
\end{center}
\end{table}

\setcounter{footnote}{5}
\renewcommand{\thefootnote}{\arabic{footnote}}

\begin{table}[t]
\caption{Trace of the transition matrices for the healthy control, ADHD-i, and ADHD-c HMM with 20 hidden states, when trained using the entire dataset of fMRIs that were reduced in dimensionality using kernel PCA.}
\label{tab:transtrace}
\begin{center}
\begin{sc}
\begin{tabular}{c c}
\hline
HMM & Transition Matrix Trace \\
\hline
Healthy & 1.9496 \\
ADHD-i & 1.9490 \\
ADHD-c & 1.9484 \\
\hline
\end{tabular}
\end{sc}
\end{center}
\end{table}

Although the results of our proposed temporal ADHD classification model are comparable to state-of-the-art ADHD classifiers published in the literature \cite{caffo2012}, our system barely performs above the threshold of labeling every input data instance as the majority class in the training dataset. The majority class is the \emph{healthy} subject type and amounts to 62.13\% of the dataset. This is compared to our best result of 63.01\% when using 20 hidden states and principal components analysis for dimensionality reduction of the fMRI.

There are several explanations for the lackluster performance of the proposed classifier, some of which pertain to the performance of ADHD classification systems in general. Perhaps the difference between the resting state brain activity for individuals who are ADHD positive and ADHD negative is negligible. It could also be the case that the regions of interest extracted from the resting-state fMRI do not contain discriminative information for diagnosing ADHD and other regions of interest should be explored. Another possible explanation is that the sampling frequency of the physical fMRI machine does not coincide with the frequency with which the brain switches mental processes and our temporal model is not synchronized with the real mental state transitions of the brain.

It could also be that the dimensionality reduction algorithms that we explored in this paper severely degrade the discriminative power of the raw voxel intensities. On the other hand, it could be that the dimensionality reduction methods still yield too many features and feature selection algorithms should be applied to the observations posed to the HMMs. To see if this is the case, we performed pairwise t-tests on the voxel mean features for the twelve regions of interest for ADHD and found that region number 4 (\emph{Intra-Calcarine Cortex}), 5 (\emph{Frontal Medial Cortex}), 9 (\emph{Cingulate Gyrus, posterior division}), and 10 (\emph{Frontal Orbital Cortex}) do not exhibit any discriminative power between healthy controls and ADHD positive subjects. That is, the null hypothesis that a set of features are not important is true for these regions of interest.

\section{Conclusion}
\label{sec:conclusion}

The development of automatic ADHD diagnostic algorithms from fMRI data is a challenging task. The application of statistical pattern recognition algorithms to this problem currently yield insubstantial results, rendering these classification systems unfit for practice in the health-care industry. However, much research is being done to improve these results and search for discriminative features for classifying ADHD amongst the plethora of voxel values present in a single fMRI scan.

Apart from systems that aggregate fMR images over time to produce a single three-dimensional image of the brain that is then used for classification, in this paper we explored a temporal classification system that uses the evolution of voxel values over time to make decisions regarding ADHD diagnosis. Specifically, we used the ADHD-200 competition dataset to learn an HMM classifier that classifies the subject as healthy, ADHD inattentive type, or ADHD combined type from an input fMRI scan. We investigated and evaluated the application of several dimensionality reduction algorithms to twelve ADHD regions of interest in the fMRI scan. Our results indicate that a hidden Markov model with 20 hidden states processing fMRI data, where the dimensionality is reduced by the principal components analysis algorithm, yields the best results with 63.01\% 5-fold cross-validation accuracy.

Still, there is much work to be done in this area. Our analysis of the transition matrices of the trained hidden Markov models indicate that other temporal models, such as recurrent neural networks or hidden Markov models with more arcs between nodes, should be explored in future work. Moreover, we propose that locating lower dimensional sets of fMRI features that retain discriminative power for ADHD classification is the heart of the problem and that the majority of future work should focus on this task.

\small {

}

\end{document}